\documentclass[runningheads]{llncs}

\usepackage[utf8]{inputenc}
\usepackage[T1]{fontenc}
\usepackage{physics}
\usepackage{subcaption}
\usepackage{tikz,float}
\usetikzlibrary{quantikz2}
\usepackage[colorlinks]{hyperref}

\newcommand{\CNOT}{\mathsf{CNOT}}
\newcommand{\QFT}{\mathsf{QFT}}
\renewcommand{\H}{\mathsf{H}}
\newcommand{\T}{\mathsf{T}}
\newcommand{\U}{\mathsf{U}}
\newcommand{\X}{\mathsf{X}}

\newcommand{\ie}{i.e.}
\newcommand{\etal}{et \; al.}

\title{Optimizing $\T$ and $\CNOT$ Gates in Quantum Ripple-Carry Adders and Comparators}
\titlerunning{Optimizing $\T$ and $\CNOT$ Gates in QRC Arithmetic}
\author{Maxime Remaud}
\institute{Eviden Quantum Lab\\ first.last@eviden.com}

\begin{document}

	\maketitle

\begin{abstract}
	The state of the art of quantum circuits using the ripple-carry strategy for the addition and comparison of two $n$-bit numbers is presented, as well as optimizations in the Clifford+$\T$ gate set, both in terms of $\CNOT$-depth and $\T$-depth, or $\CNOT$-count and $\T$-count. In particular, we consider the adders presented by Cuccaro $\etal$ and Takahashi $\etal$, and exhibit an adder with a $\T$-depth of $3n$ and a $\CNOT$-depth of $8n$, while without optimization of the original circuits, a $\T$-depth of $6n$ is expected. Note that we have focused here on quantum ripple-carry adders using at most one ancilla, without any approximation of the 3-qubit gates involved (Toffoli, Peres and TR) or any strategy involving a measurement.
\end{abstract}

\section{Introduction}

When considering the complexity of implementing a circuit on a quantum computer, several questions arise, not the least of which is which gates we can use for our purpose. To answer this question, it is usual to consider a decomposition of the circuit into 1- or 2-qubit gates, for the purpose of comparing different techniques, but also for practical reasons: quantum computers can only use these elementary gates. It is then necessary to choose a set of gates into which any other gate could be decomposed, and to define metrics to evaluate the performance of an implementation.

\paragraph{Universal gate sets.}

It is well known that the set of all the 1-qubit and 2-qubit gates is {\em universal}, meaning that it can be used to decompose any quantum gate acting on any number of qubits \cite{KLM06}. However, it is not reasonable to assume that a quantum computer capable of executing any gate of this universal gate set could ever be built, because of the infinite number of gates this set contains. For this reason, we usually consider gate sets that are {\em approximately} universal, meaning that they typically contain a limited number of gates that are sufficient to approximate any quantum gate with any desired precision \cite{Chi22}. In this paper, we will consider the set Clifford$+\T$, which has been widely studied because of its potential use in fault-tolerant quantum circuit design.

\paragraph{Metrics.}

When it comes to assessing the complexity of a quantum circuit, we have three main metrics at our disposal. The first is {\em width}, which corresponds to the number of qubits required to implement the circuit in question and gives an indication of the memory required to run the circuit. In this paper we will only look at circuits that use at most one ancillary qubit, so the width is not relevant for comparison. The second and third metrics are what we will call the {\em size} and the {\em depth} of the circuit. They correspond respectively to the number of gates and the number of time slices in the circuit and, roughly speaking, give an indication of how long the circuit will take to do its job, $\ie$, on the computation time. 

In particular, when considering the Clifford+$\T$ gate set, it is well known that the $\T$ gate is the most expensive one of the set. A number of studies have naturally tried to optimize its use, both in terms of depth and count. However, minimizing the depth or count of $\T$ gates in a circuit has the side effect of increasing the count or depth of $\CNOT$ gates. Since the $\CNOT$ gate is a 2-qubit gate, we cannot afford to ignore this additional cost. Thus, it is important to analyze the complexity of our circuits in terms of both $\T$-count and $\T$-depth, as well as $\CNOT$-count and $\CNOT$-depth. For further discussion, please refer to \cite{Mas16,BBVMA}.

\paragraph{Gates.}

\begin{figure}[H]\centering
	\subcaptionbox{\label{circ:PeresSymb}}{\scalebox{0.8}{
	\begin{quantikz}[column sep=.2cm, row sep={.6cm,between origins}]
		& \ctrl{1} & \\
		& \ctrlo{1} & \\
		& \targ{} &
	\end{quantikz}}}
	\subcaptionbox{\label{circ:TRSymb}}{\scalebox{0.8}{
	\begin{quantikz}[column sep=.2cm, row sep={.6cm,between origins}]
		& \ctrl{1} & \\
		& \ctrlo{1} & \\
		& \targo{} &
	\end{quantikz}}}
	\hspace{.2cm}
	\subcaptionbox{\label{circ:Toffoli}}{\scalebox{0.8}{
	\begin{quantikz}[column sep=.2cm, row sep={.6cm,between origins}]
		& \ctrl{1}	& \push{\quad \textcolor{white}{=} \quad}& \ctrl{1} 	& \ctrl{1} 	& \push{\quad \textcolor{white}{=} \quad}& \ctrl{1}	& \ctrl{1} & \\
		& \ctrl{1}	& \push{\quad = \quad}		& \ctrlo{1}	& \targ{}  	& \push{\quad = \quad}		& \targ{}	& \ctrlo{1}& \\
		& \targ{} 	& \push{\quad \textcolor{white}{=} \quad}& \targ{} 	& 			& \push{\quad \textcolor{white}{=} \quad}& 			& \targo{} &			
	\end{quantikz}}}
	\caption{(a) Symbol for the Peres gate. (b) Symbol for the TR gate. (c) Symbol for the Toffoli gate and relations with the previous gates.}\label{circ:Gates}
\end{figure}
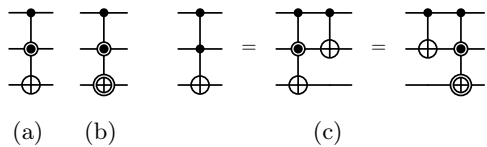

In order to design arithmetic circuits, some 3-qubit gates are particularly useful, namely the well-known Toffoli gate, Peres gate and TR gate (which is the reversed Peres gate). See \autoref{circ:Gates} for their circuit representation. Several works showed how to implement these gates using the Clifford+$\T$ gate set \cite{TR10,AMMR13,KMR14}.

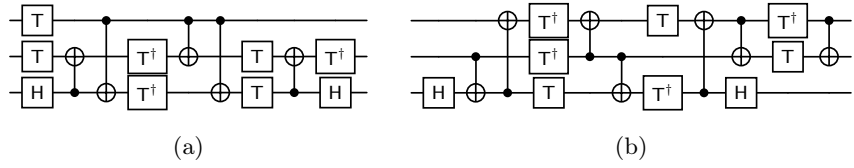
\begin{figure}[H]\centering
	\subcaptionbox{\label{circ:PeresDec}}{\scalebox{0.8}{\begin{quantikz}[column sep=.2cm, row sep={.6cm,between origins}]
		& \gate{\T}	& 			& \ctrl{2}	& 			 	& \ctrl{1} 	& \ctrl{2}	&			&			& 					& \\
		& \gate{\T} & \targ{} 	& 			& \gate{\T^\dag}& \targ{} 	& 			& \gate{\T} & \targ{}	& \gate{\T^\dag}	& \\
		& \gate{\H} & \ctrl{-1} & \targ{}	& \gate{\T^\dag}& 			& \targ{}	& \gate{\T} & \ctrl{-1} & \gate{\H}			&
	\end{quantikz}}}
	\hspace{.2cm}
	\subcaptionbox{\label{circ:ToffDec}}{\scalebox{0.8}{\begin{quantikz}[column sep=.2cm, row sep={.6cm,between origins}]
		& 			& 			& \targ{}	& \gate{\T^\dag}& \targ{} 	&			& \gate{\T} 	& \targ{} 	& \ctrl{1} 	& \gate{\T^\dag}& \ctrl{1} 	& \\
		& 			& \ctrl{1}	&			& \gate{\T^\dag}& \ctrl{-1} & \ctrl{1} 	& 				&			& \targ{} 	& \gate{\T} 	& \targ{} 	& \\
		& \gate{\H} & \targ{}	& \ctrl{-2} & \gate{\T}		& 			& \targ{} 	& \gate{\T^\dag}& \ctrl{-2} & \gate{\H} &				&			&
	\end{quantikz}}}
	\caption{(a) Decomposition of the Peres gate with 5 $\CNOT$ gates but a $\T$-depth of 4. (b) Decomposition of the Toffoli gate with a $\T$-depth of 3 and 7 $\CNOT$ gates.}\label{circ:GatesDec}
\end{figure}

Even when we look at the decomposition of the Peres gate (or the Toffoli or TR gate) in the Clifford+$\T$ gate set, we may have to choose between two possibilities: either implement it by optimizing the $\CNOT$-count, at the expense of having a somewhat high $\T$-depth, or minimize the latter, at the expense of increasing the $\CNOT$-count. See \autoref{circ:GatesDec} for the circuits used in this paper, taken from \cite{AMMR13}.

\subsection{Ripple-carry technique}

The aim of this paper is to study in detail how addition and comparison (which are basic arithmetic operations, essential to the realization of many algorithms) can be achieved on a quantum computer. In particular, we focus on the ripple-carry technique that we present here.

We will work with two $n$-bit numbers denoted $a$ and $b$. We write their respective binary expansion $(a_{n-1},\dots,a_0)$ and $(b_{n-1},\dots,b_0)$, $a_0$ and $b_0$ being the least significant bits. 

\paragraph{Comparison.}

The action of a comparator is described as follows:
\begin{equation*}
	\ket{a} \ket{b} \ket{z} \mapsto \ket{a} \ket{b} \ket{z \oplus y}
\end{equation*}
where $y=1$ if and only if $a \le b$ ($\ie$, $y=0$ if and only if $a > b$).

But when coming to the task of determining if $a$ is greater than $b$, equal to $b$, or smaller than $b$, it actually all comes down to the task of adding $a$ and $b$. Indeed, the boolean equal to $(a > b)$ is directly linked to the most significant bit of $a+b$ \cite{CDKM04}. We will not go in further details here: since comparison reduces to addition, we focus on addition.

\paragraph{Addition.}

We compute the addition in place, meaning that we want to build an algorithm with the following action:
\begin{equation*}
	\ket{a}_n \ket{b}_n \ket{z} \mapsto \ket{a}_n \ket{a+b}_n \ket{z \oplus (a+b)_n}.
\end{equation*}
The register initially containing a bit $z$ will be overloaded with the bit of overflow of the sum. In addition, note that our circuit should be reversible, $\ie$, if we have to use ancillary qubits, they have to be reset to zero at the end of the circuit.

If we call $s$ the sum of $a$ and $b$ (we will write $(s_{n},\dots,s_0)$ its binary expansion) and $c$ the string of successive carries encountered during the addition process, we have the following recursive definitions:
\begin{equation}\label{eq:DefCarry}
	c_i = \begin{cases} 
		0 & \text{if} \; i=0 \\
		a_{i-1} b_{i-1} \oplus b_{i-1} c_{i-1} \oplus c_{i-1} a_{i-1} & \text{for} \; i \in [\![1,n]\!]
	\end{cases}
\end{equation}
and
\begin{equation}\label{eq:DefSum}
	s_i = \begin{cases}
		a_i \oplus b_i \oplus c_i & \text{for} \; i \in [\![0,n-1]\!] \\
		c_n & \text{if} \; i=n.
	\end{cases}
\end{equation}

It is possible to use the recursive definition of Equation \eqref{eq:DefCarry} to calculate all the carries recursively, from their predecessor and the corresponding bits of $a$ and $b$, which forms a cascade when viewed as a circuit. When all the carries have been obtained, the bits of $s$ are computed thanks to Equation \eqref{eq:DefSum} and the carries uncomputed at the same time, this time forming a mirrored cascade to the previous one. This process is known as {\em ripple-carry addition} and easily recognizable by its V-shaped circuit. A non-exhaustive review of quantum ripple-carry adders can be found for example in \cite{OOCG20}.
For the rest of this paper, we focus on adders using only $O(1)$ ancillary qubits. We will see that such quantum ripple-carry adders with $O(n)$ size and $O(n)$ depth can be designed. Our goal will be to reduce the prefactors hidden in the big O notation, by dissecting previously known and new circuits in the Clifford+$\T$ gate set and optimizing depth and count of $\CNOT$ and $\T$ gates.

\paragraph{Remark.}

A second way of computing the sum is known as {\em carry-lookahead} addition. Adders of this kind have the advantage of having a depth of $O(\log{n})$ but the main drawback of requiring $O(n)$ ancillary qubits \cite{DKR06,TK08,TJNA13,Mog19}. A third and last method which exploits the quantum Fourier transform has the advantage to not use any ancillary qubit but the drawback of a size of $O(n^2)$ \cite{Dra02,RPGE17}.

All in all, quantum ripple-carry adders and adders using the $\QFT$ are both space-efficient, but the former has the additional advantage of being suitable for LNN architectures. Thus, we focus in this paper on using the ripple-carry technique to design new more efficient quantum arithmetic circuits.

\subsection{In this paper}

In the second section of this paper, we give several optimization rules, $\ie$, equivalences between circuits involving 3-qubit gates and circuits optimized in the Clifford+$\T$ gate set. These optimization rules will be useful for analyzing the complexity in terms of depth and count in $\T$ and $\CNOT$ gates of circuits written using Toffoli, Peres and TR gates.

In \autoref{sec:Shallow}, we apply some of these rules on one of Cuccaro $\etal$ adders \cite{CDKM04} to refine its complexity in the Clifford+$\T$ gate set. It turns out that it is the most interesting in terms of depth, since we show that it has a $\T$-depth of $3n+O(1)$ and a $\CNOT$-depth of $8n+O(1)$. Then, in \autoref{sec:Cheap}, we optimize the other adder introduced by Cuccaro $\etal$, which is the most interesting in terms of gates count: it has a $\CNOT$-count of $14n-O(1)$ and a $\T$-count of $10n-O(1)$. Finally, in \autoref{sec:AncFree}, we take back Takahashi $\etal$ \cite{TTK10} adder and proceed to optimize it. It yields an ancilla-free adder that has also a $\T$-depth of $3n+O(1)$.

In the last section, we look at comparators. Exactly as for the adders, we take up the works of \cite{CDKM04} and \cite{TTK10} and apply our optimization rules to them, thus obtaining new circuits with more interesting metrics than without dissecting the 3-qubit gates.

A summary of the complexity of the various known and new ripple-carry adders using at most one ancillary qubits is given in \autoref{tab:RCAdd}, while \autoref{table:Comp} gives the same for the comparators. 

\begin{table}[h!]
	\centering
	\begin{tabular}{|c||c|c|c|c|c|}
	\hline
	Algo.  		    & $\CNOT$- 	& $\CNOT$-	& $\T$-	& $\T$-	& Ancilla  \\
	                & depth	   & count	& depth	& count	&   \\
	\hline\hline
	\cite{TK05}		& $26n-42$		& $34n-41$ 		& $9n-9$		& $28n-35$		& $0$	\\
	\cite{SRV08}		& $16n+3$		& $18n+1$		& $6n$			& $14n$			& $1$ 	\\
	\cite{CDKM04} 	& $16n-25$		& $18n-18$		& $6n-9$		& $14n-21$		& $1$   \\
	\cite{TTK10}		& $15n-8$		& $17n-12$		& $6n-3$		& $14n-7$		& $0$	\\
	\cite{TR11}		& $14n-1$		& $18n-6$		& $6n-3$		& $14n-7$		& $1$	\\
	\cite{CDKM04} 	& $13n-3$		& $17n-10$ 		& $6n-3$		& $14n-7$		& $1$ 	\\
	\autoref{sec:Cheap} & $11n-8$	& $\mathbf{14n-10}$	& $4n-2$	& $\mathbf{10n-3}$	& $1$ \\
	\autoref{sec:AncFree} &	$10n-3$	& $16n-12$		& $\mathbf{3n+2}$& $12n-5$		& $\mathbf{0}$ \\
	\autoref{sec:Shallow}& $\mathbf{8n+2}$	& $16n-10$	& $\mathbf{3n+2}$	& $12n-5$	& $1$ \\
	\hline
	\end{tabular}
	\caption{Complexity of ripple-carry adders (using only 0 or 1 ancillary qubit) ordered by decreasing $\CNOT$-depth.}\label{tab:RCAdd}
\end{table}

We have not included the adder proposed in \cite{LFXP20} in this state-of-the-art, as it can be verified that it does not implement what is claimed. This is due to a poor definition of what is called the TR2 gate (see their Figure 4(c)). This gate is supposed to map $\ket{C,B,A}$ to $\ket{A\bar{B} \oplus C, B, A \oplus B}$ when in fact it maps $\ket{C,B,A}$ to $\ket{\bar{A}B \oplus C, B, A \oplus B}$. The relationship between the Peres gate "PG1" and this TR gate "TR2" is therefore also incorrect (see their Figure 7(c)). Meanwhile, it is the key element in the construction used to build the proposed adder, which is in turn flawed.

\begin{table}[h!]
	\centering
	\begin{tabular}{|c||c|c|c|c|c|}
	\hline
	Algo.  		    & $\CNOT$- 	& $\CNOT$-	& $\T$-	& $\T$-	& Ancilla  \\
	                & depth	   & count	& depth	& count	&   \\
	\hline\hline
	\cite{TA09} 		& $18n+3$		& $20n+1$		& $6n$			& $14n$			& $1$		 \\
	\cite{TTK10}		& $16n-8$		& $18n-12$		& $6n-3$		& $14n-7$		& $0$ \\
	\cite{CDKM04}	& $14n-3$		& $18n-7$		& $6n-3$		& $14n-7$		& $1$ \\
	\autoref{sec:CompTTK}& $10n-6$	& $\mathbf{14n-9}$	& $\mathbf{4n-3}$	& $\mathbf{10n-6}$		& $\mathbf{0}$ \\
	\autoref{sec:CompCDKM}& $\mathbf{8n+5}$& $\mathbf{14n-6}$	& $\mathbf{4n-1}$& $\mathbf{10n-3}$	& $1$ \\
	\hline
	\end{tabular}
	\caption{Complexity of comparators (using only 0 or 1 ancillary qubit) built from the ripple-carry strategy, ordered by decreasing $\CNOT$-depth.}\label{table:Comp}
\end{table}

We note that the comparator described in \cite{XLZ18} corresponds to the comparator derived from \cite{CDKM04}, the one described in \cite{LFXP20} is equivalent to the comparator derived from \cite{TTK10}, and that the comparators described in \cite{Mog18} and \cite{XZS20} are equivalent to the comparator described in \cite{TA09} (derived from \cite{SRV08}), reason why they do not appear in \autoref{table:Comp}.

\section{Preliminaries: Optimization Rules}

In this section, we draw inspiration from the work of \cite{LFXL22}, where circuit optimizations are performed using different “optimization rules”. The difference between Li $\etal$ work and ours lies in the use of approximate (Toffoli, Peres and TR) gates. One of the results of their paper is to show that with phase approximations, designing an adder of $\T$-depth $2n$ is possible. We show that we can in fact design circuits of $\T$-depth of the order of $4n$ without any phase approximation, which is better than any previous work on quantum ripple-carry addition without measurement nor approximation, using the appropriate optimization rules. We give hereafter the rules we will use to optimize the adders and comparators studied in the rest of this paper.

\subsection{Toffoli-Peres V-shaped circuit}\label{sec:ToffPer}

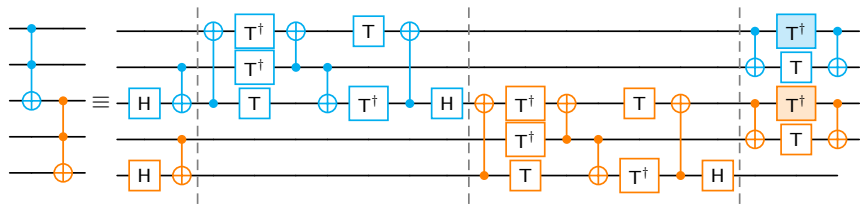
\begin{figure}[H]\centering
	\scalebox{0.8}{
	\begin{quantikz}[column sep=.2cm, row sep={.6cm,between origins}]
		& \ctrl[style={draw=cyan,fill=cyan}]{1} & & \\
		& \ctrl[style={draw=cyan,fill=cyan}]{1} & & \\
		& \targ[style={draw=cyan}]{} & \ctrl[style={draw=orange,fill=orange}]{1} & \\
		& & \ctrl[style={draw=orange,fill=orange}]{1} & \\
		& & \targ[style={draw=orange}]{} & 
	\end{quantikz}}$\equiv$\scalebox{0.8}{\begin{quantikz}[column sep=.2cm, row sep={.6cm,between origins}]
		& 			& \slice[style={draw=gray}]{}	& \targ[style={draw=cyan}]{}	& \gate[style={draw=cyan}]{\T^\dag}& \targ[style={draw=cyan}]{} 	&   & \gate[style={draw=cyan}]{\T} & \targ[style={draw=cyan}]{} 	& 								\slice[style={draw=gray}]{} & &		&				&			&			& & \slice[style={draw=gray}]{}		& \ctrl[style={draw=cyan,fill=cyan}]{1} 	& \gate[style={draw=cyan,fill=cyan!20}]{\T^\dag} 	& \ctrl[style={draw=cyan,fill=cyan}]{1}	& \\
		& 			& \ctrl[style={draw=cyan,fill=cyan}]{1}	&			& \gate[style={draw=cyan}]{\T^\dag}& \ctrl[style={draw=cyan,fill=cyan}]{-1} & \ctrl[style={draw=cyan,fill=cyan}]{1} 	& 		&	&					& 			&			&				&			&			&		&	& \targ[style={draw=cyan}]{} 	& \gate[style={draw=cyan}]{\T} 		& \targ[style={draw=cyan}]{}	& \\
		& \gate[style={draw=cyan}]{\H} & \targ[style={draw=cyan}]{}	& \ctrl[style={draw=cyan,fill=cyan}]{-2} & \gate[style={draw=cyan}]{\T}		& 			& \targ[style={draw=cyan}]{} 	& \gate[style={draw=cyan}]{\T^\dag} 	& \ctrl[style={draw=cyan,fill=cyan}]{-2} & \gate[style={draw=cyan}]{\H} & \targ[style={draw=orange}]{}	& \gate[style={draw=orange}]{\T^\dag}& \targ[style={draw=orange}]{} 	&   & \gate[style={draw=orange}]{\T} & \targ[style={draw=orange}]{}	& & \ctrl[style={draw=orange,fill=orange}]{1} 	& \gate[style={draw=orange,fill=orange!20}]{\T^\dag} 	& \ctrl[style={draw=orange,fill=orange}]{1}	& \\
		& 			& \ctrl[style={draw=orange,fill=orange}]{1}	&			&				&			&			&			&					&			&			& \gate[style={draw=orange}]{\T^\dag}& \ctrl[style={draw=orange,fill=orange}]{-1} & \ctrl[style={draw=orange,fill=orange}]{1} 	& 			&  &  & \targ[style={draw=orange}]{} 	& \gate[style={draw=orange}]{\T} 		& \targ[style={draw=orange}]{}	& \\
		& \gate[style={draw=orange}]{\H} & \targ[style={draw=orange}]{}	&			&				&			&			&			&					&			& \ctrl[style={draw=orange,fill=orange}]{-2} & \gate[style={draw=orange}]{\T}		& 			& \targ[style={draw=orange}]{} 	& \gate[style={draw=orange}]{\T^\dag} 	& \ctrl[style={draw=orange,fill=orange}]{-2}& \gate[style={draw=orange}]{\H} & & 			&
	\end{quantikz}}
	\caption{Left cascade of two Toffoli gates. Its $\T$-depth is 5, $\T$-count is 14, $\CNOT$-depth is 11, and $\CNOT$-count is 14.}\label{circ:ToffoliLeft}
\end{figure}

\begin{figure}[H]\centering
	\scalebox{0.8}{
	\begin{quantikz}[column sep=.2cm, row sep={.6cm,between origins}]
		& & \ctrl[style={draw=green,fill=green}]{1} & \\
		& & \ctrlo[style={draw=green,fill=green}]{1} & \\
		& \ctrl[style={draw=magenta,fill=magenta}]{1} & \targ[style={draw=green}]{} & \\
		& \ctrlo[style={draw=magenta,fill=magenta}]{1} & & \\
		& \targ[style={draw=magenta}]{} & & 
	\end{quantikz}}$\equiv$\scalebox{0.8}{\begin{quantikz}[column sep=.2cm, row sep={.6cm,between origins}]
		& \gate[style={fill=green!20,draw=green}]{\T}	\slice[style={draw=gray}]{} &			&			&				&			& \slice[style={draw=gray}]{}	&			& 			& \ctrl[style={draw=green,fill=green}]{2}	& 			 	& \ctrl[style={draw=green,fill=green}]{1} 	& \ctrl[style={draw=green,fill=green}]{2}	 \slice[style={draw=gray}]{} &			&			& 					& \\
		& \gate[style={draw=green}]{\T}	&			&			&				&			&			&			& \targ[style={draw=green}]{} 	& 			& \gate[style={draw=green}]{\T^\dag}& \targ[style={draw=green}]{} 	& 			& \gate[style={draw=green}]{\T} & \targ[style={draw=green}]{}	& \gate[style={draw=green}]{\T^\dag}	& \\
		& \gate[style={draw=magenta,fill=magenta!20}]{\T}	& 			& \ctrl[style={draw=magenta,fill=magenta}]{2}	& 			 	& \ctrl[style={draw=magenta,fill=magenta}]{1} 	& \ctrl[style={draw=magenta,fill=magenta}]{2}	& \gate[style={draw=green}]{\H} & \ctrl[style={draw=green,fill=green}]{-1} & \targ[style={draw=green}]{}	& \gate[style={draw=green}]{\T^\dag}& 			& \targ[style={draw=green}]{}	& \gate[style={draw=green}]{\T} & \ctrl[style={draw=green,fill=green}]{-1} & \gate[style={draw=green}]{\H}			& \\
		& \gate[style={draw=magenta}]{\T} & \targ[style={draw=magenta}]{} 	& 			& \gate[style={draw=magenta}]{\T^\dag}& \targ[style={draw=magenta}]{} 	& 			&			&			&			& 				&			&			& \gate[style={draw=magenta}]{\T} & \targ[style={draw=magenta}]{}	& \gate[style={draw=magenta}]{\T^\dag}	& \\
		& \gate[style={draw=magenta}]{\H} & \ctrl[style={draw=magenta,fill=magenta}]{-1} & \targ[style={draw=magenta}]{}	& \gate[style={draw=magenta}]{\T^\dag}& 			& \targ[style={draw=magenta}]{}	& 			&			&			&				&			&			& \gate[style={draw=magenta}]{\T} & \ctrl[style={draw=magenta,fill=magenta}]{-1} & \gate[style={draw=magenta}]{\H}			&
	\end{quantikz}}
	\caption{Right cascade of two Peres gates. Its $\T$-depth is 5, $\T$-count is 14, $\CNOT$-depth is 9, and $\CNOT$-count is 10.}\label{circ:PeresRight}
\end{figure}
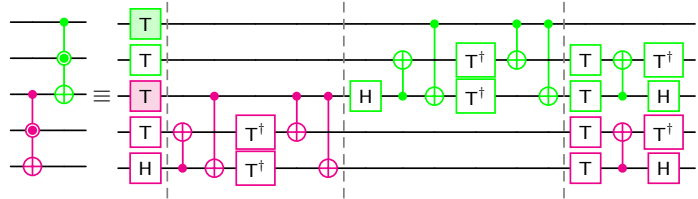

In the last two sections, we will look at circuits involving a cascade of Toffoli gates on the left-hand branch and of Peres gates on the right-hand branch. In \autoref{circ:ToffoliLeft}, we give the decomposition in the Clifford+$\T$ gate set of the left cascade of Toffoli gates and in \autoref{circ:PeresRight} the decomposition of the right cascade of Peres gates.

From \autoref{circ:ToffoliLeft}, we easily generalize that for a left-hand cascade of $n$ Toffoli gates, the $\T$-depth is $2n+1$, $\T$-count is $7n$, $\CNOT$-depth is $4n+3$, and $\CNOT$-count is $7n$.

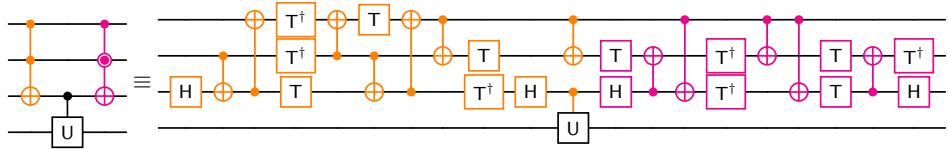
\begin{figure}[H]\centering
	\scalebox{0.8}{
	\begin{quantikz}[column sep=.2cm, row sep={.6cm,between origins}]
		& \ctrl[style={draw=orange,fill=orange}]{1} & 		 & \ctrl[style={draw=magenta,fill=magenta}]{1} & \\
		& \ctrl[style={draw=orange,fill=orange}]{1} & 		 & \ctrlo[style={draw=magenta,fill=magenta}]{1} & \\
		& \targ[style={draw=orange}]{} & \ctrl{1} & \targ[style={draw=magenta}]{} & \\
		& & \gate{\U} & & 
	\end{quantikz}}$\equiv$\scalebox{0.8}{\begin{quantikz}[column sep=.2cm, row sep={.6cm,between origins}]
		& 			& 			& \targ[style={draw=orange}]{}	& \gate[style={draw=orange}]{\T^\dag}& \targ[style={draw=orange}]{} 	& \gate[style={draw=orange}]{\T} & \targ[style={draw=orange}]{} 	& \ctrl[style={draw=orange,fill=orange}]{1} 	& 					& 			& \ctrl[style={draw=orange,fill=orange}]{1} &			& 			& \ctrl[style={draw=magenta,fill=magenta}]{2}	& 			 	& \ctrl[style={draw=magenta,fill=magenta}]{1} 	& \ctrl[style={draw=magenta,fill=magenta}]{2}	&			&			& 					& \\
		& 			& \ctrl[style={draw=orange,fill=orange}]{1}	&			& \gate[style={draw=orange}]{\T^\dag}& \ctrl[style={draw=orange,fill=orange}]{-1} & \ctrl[style={draw=orange,fill=orange}]{1} 	& 			& \targ[style={draw=orange}]{} 	& \gate[style={draw=orange}]{\T} 		& 			& \targ[style={draw=orange}]{} & \gate[style={draw=magenta}]{\T} & \targ[style={draw=magenta}]{} 	& 			& \gate[style={draw=magenta}]{\T^\dag}& \targ[style={draw=magenta}]{} 	& 			& \gate[style={draw=magenta}]{\T} & \targ[style={draw=magenta}]{}	& \gate[style={draw=magenta}]{\T^\dag}	& \\
		& \gate[style={draw=orange}]{\H} & \targ[style={draw=orange}]{}	& \ctrl[style={draw=orange,fill=orange}]{-2} & \gate[style={draw=orange}]{\T}		& 			& \targ[style={draw=orange}]{} 	& \ctrl[style={draw=orange,fill=orange}]{-2} & 			& \gate[style={draw=orange}]{\T^\dag} 	& \gate[style={draw=orange}]{\H} & \ctrl[style={draw=orange,fill=orange}]{1} & \gate[style={draw=magenta}]{\H} & \ctrl[style={draw=magenta,fill=magenta}]{-1} & \targ[style={draw=magenta}]{}	& \gate[style={draw=magenta}]{\T^\dag}& 			& \targ[style={draw=magenta}]{}	& \gate[style={draw=magenta}]{\T} & \ctrl[style={draw=magenta,fill=magenta}]{-1} & \gate[style={draw=magenta}]{\H}			& \\
		&&&&&&&&&&& \gate{\U} &&&&&&&&&& \\
	\end{quantikz}}
	\caption{V-shaped circuit with Toffoli gates on the left-hand branch and Peres gates on the right-hand one. The color filled gates from \autoref{circ:ToffoliLeft} and \autoref{circ:PeresRight} cancel.}\label{circ:ToffPeres}
\end{figure}

From \autoref{circ:PeresRight}, we generalize that for a right-hand cascade of $n$ Peres gates, the $\T$-depth is $n+3$, $\T$-count is $7n$, $\CNOT$-depth is $4n+1$, and $\CNOT$-count is $5n$. In \autoref{circ:ToffPeres} we look at the V-shaped construction with Toffoli gates on the left-hand branch and Peres gates on the right-hand one.

If we inject \autoref{circ:ToffoliLeft} and \autoref{circ:PeresRight} in \autoref{circ:ToffPeres}, we obtain for a V-shaped circuit with $n$ layers: a $\T$-depth of $3n+4$, a $\T$-count of $14n$, a $\CNOT$-depth of $8n+4$ and a $\CNOT$-count of $12n$. Simplifying the matching pairs of $\T$ and $\T^\dag$ gates as in the present circuit, layer by layer, we can reduce the $\T$-count by $2n$.

\subsection{Toffoli-Toffoli V-shaped circuit}\label{sec:ToffToff}

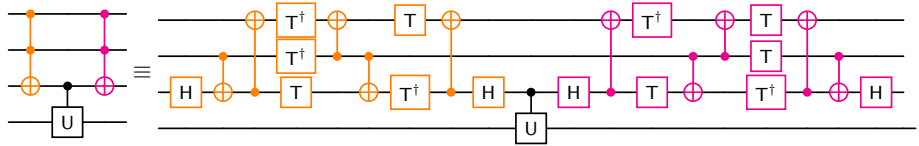
\begin{figure}[H]\centering
	\scalebox{0.8}{
	\begin{quantikz}[column sep=.2cm, row sep={.6cm,between origins}]
		& \ctrl[style={draw=orange,fill=orange}]{1} 	& 			& \ctrl[style={draw=magenta,fill=magenta}]{1} 	& \\
		& \ctrl[style={draw=orange,fill=orange}]{1} & 			& \ctrl[style={draw=magenta,fill=magenta}]{1}	& \\
		& \targ[style={draw=orange}]{} 	& \ctrl{1} 	& \targ[style={draw=magenta}]{} 	& \\
		& 			& \gate{\U} &			&
	\end{quantikz}}$\equiv$\scalebox{0.8}{\begin{quantikz}[column sep=.2cm, row sep={.6cm,between origins}]
		&			& 			& \targ[style={draw=orange}]{}	& \gate[style={draw=orange}]{\T^\dag} 	& \targ[style={draw=orange}]{} 	& & \gate[style={draw=orange}]{\T}	& \targ[style={draw=orange}]{} &	 & & & \targ[style={draw=magenta}]{}& \gate[style={draw=magenta}]{\T^\dag} & 	& \targ[style={draw=magenta}]{}	& \gate[style={draw=magenta}]{\T} & \targ[style={draw=magenta}]{}	&			& 				& \\
		& 			& \ctrl[style={draw=orange,fill=orange}]{1} 	& 			& \gate[style={draw=orange}]{\T^\dag}& \ctrl[style={draw=orange,fill=orange}]{-1} 	& \ctrl[style={draw=orange,fill=orange}]{1}	 & & & & & & & & \ctrl[style={draw=magenta,fill=magenta}]{1}& \ctrl[style={draw=magenta,fill=magenta}]{-1}  & \gate[style={draw=magenta}]{\T} & & \ctrl[style={draw=magenta,fill=magenta}]{1}	& & \\
		& \gate[style={draw=orange}]{\H} & \targ[style={draw=orange}]{} & \ctrl[style={draw=orange,fill=orange}]{-2}	& \gate[style={draw=orange}]{\T}& 			& \targ[style={draw=orange}]{}	& \gate[style={draw=orange}]{\T^\dag} & \ctrl[style={draw=orange,fill=orange}]{-2} & \gate[style={draw=orange}]{\H}	& \ctrl{1}	& \gate[style={draw=magenta}]{\H}	& \ctrl[style={draw=magenta,fill=magenta}]{-2} & \gate[style={draw=magenta}]{\T}& \targ[style={draw=magenta}]{}	& 			& \gate[style={draw=magenta}]{\T^\dag}		& \ctrl[style={draw=magenta,fill=magenta}]{-2}	& \targ[style={draw=magenta}]{-1} & \gate[style={draw=magenta}]{\H} 	& \\
		&			&			&			&				&			&			&			&			&			& \gate{\U} &&&&&&&&&&&
	\end{quantikz}}
	\caption{V-shaped circuit with Toffoli gates on both branches. The last slice of \autoref{circ:ToffoliLeft} and its reverse match and cancel.}\label{circ:ToffToff}
\end{figure}

In the last two sections, we will look at V-shaped circuits with Toffoli gates on both branches. In \autoref{circ:ToffoliLeft}, we already gave the decomposition in the Clifford+$\T$ gate set of the left cascade of Toffoli gates. For the mirrored branch, the exact same circuit in reverse is what we will use. From \autoref{sec:ToffPer}, we thus obtain the V-shaped circuit with Toffoli gates on both branches presented in \autoref{circ:ToffToff}.

If we inject \autoref{circ:ToffoliLeft} and its reversed version in \autoref{circ:ToffToff}, we obtain for a V-shaped circuit with $n$ layers: a $\T$-depth of $4n+2$, a $\T$-count of $14n$, a $\CNOT$-depth of $8n+6$ and a $\CNOT$-count of $14n$. Simplifying the matching $\T$ and $\CNOT$ gates as in the present circuit, layer by layer, we can reduce the $\T$-count by $4n$, the $\T$-depth by 2, the $\CNOT$-count by $4n$ and the $\CNOT$-depth by 4.

\section{Optimizing Ripple-Carry Adders}

In what follows, we take the two adders introduced by Cuccaro $\etal$'s in \cite{CDKM04}, one being depth-efficient and the other being size-efficient, as well as Takahashi $\etal$'s adder \cite{TTK10}, which does not need any ancilla. We optimize them thanks to the rules defined previously.

\subsection{Shallow Ripple-Carry Adder}\label{sec:Shallow}

We start with a circuit from \cite{CDKM04}, namely their Figure 6, or Figure 4 combined with Figure 2(b). An example for $n=6$ is provided in \autoref{circ:Cuccaro}. A V-shaped circuit can be find in there, with a left branch of Toffoli gates and a right branch of Peres gates.

Without any optimization, it can be verified that for any $n$, Slices 1 and 4 of their circuit have a $\CNOT$-depth of 2 and $\CNOT$-count of $n$. Slice 2 has a $\CNOT$-depth of $7(n-1)$, a $\CNOT$-count of $8(n-1)$, a $\T$-depth of $3(n-1)$ and a $\T$-count of $7(n-1)$. Slice 3 has a $\CNOT$-depth of $6n$, a $\CNOT$-count of $7n-2$, a $\T$-depth of $3n$ and a $\T$-count of $7n$. In total, a $\CNOT$-depth of $13n-3$, $\CNOT$-count of $17n-10$, $\T$-depth of $6n-3$ and $\T$-count of $14n-7$ are exepected.

Now, taking back the metrics given in \autoref{sec:ToffPer} for a V-shaped circuit with Toffoli gates on the left branch and Peres gates on the right one (with $n-1$ layers): we have for the optimized Cuccaro $\etal$'s adder a $\T$-depth of $3n+1$, a $\T$-count of $12n-12$, a $\CNOT$-depth of $8n-4$ and a $\CNOT$-count of $12n-12$. The operator $\U$ is a Peres gate. Using \autoref{circ:PeresDec}, we have to add 2 to the $\CNOT$-depth and 1 to the $\T$-depth (the majority of the gates being parallelizable with gates of the V-shape), 5 to the $\CNOT$-count and 7 to the $\T$-count. For the rest of the circuit, we have $4n-3$ additionnal $\CNOT$ gates, but the majority can once again be computed in parallel with the V-shape, thus we have an additionnal small increase of 4 for the $\CNOT$-depth. In total, the circuit in \autoref{circ:Cuccaro} has $\T$-depth of $3n+2$, $\T$-count of $12n-5$, $\CNOT$-depth of $8n+2$ and $\CNOT$-count of $16n-10$.

\begin{figure}[!ht]\centering
	\scalebox{0.8}{
	\begin{quantikz}[column sep=.2cm, row sep={.6cm,between origins}]
	\lstick{$\ket{b_0}$}	& 			& \slice{1}	& \ctrl[style={draw=orange,fill=orange}]{1} 	&			&			&			& \slice{2}	&			&			&			& 			& 			& \ctrlo[style={draw=magenta,fill=magenta}]{1} \slice{3}	& 			& \slice{4}	& \rstick{$\ket{s_0}$} \\
	\lstick{$\ket{a_0}$}	& 			& 			& \ctrl[style={draw=orange,fill=orange}]{1}	& 			& 			&			& 			&			&			&			& 			& 			& \ctrl[style={draw=magenta,fill=magenta}]{1}				& 			& 			& \rstick{$\ket{a_0}$} \\
	\lstick{$\ket{0}$}		&			& \targ{}	& \targ[style={draw=orange}]{}	& \ctrl[style={draw=orange,fill=orange}]{1}	&			&			& 			&			&			&			& 			& \ctrl[style={draw=magenta,fill=magenta}]{1}	& \targ[style={draw=magenta}]{}				& \targ{} 	& 		 	& \rstick{$\ket{0}$} \\
	\lstick{$\ket{b_1}$}	& \targ{}	& 			& 			& \ctrl[style={draw=orange,fill=orange}]{1} 	& 			&			& 			&			&			&			& 			& \ctrlo[style={draw=magenta,fill=magenta}]{1}	& 						& 			& \targ{}	& \rstick{$\ket{s_1}$} \\
	\lstick{$\ket{a_1}$}	& \ctrl{-1}	& \ctrl{-2}	& \targ{}	& \targ[style={draw=orange}]{} 	& \ctrl[style={draw=orange,fill=orange}]{1}	&			& 			&			&			&			& \ctrl[style={draw=magenta,fill=magenta}]{1}	& \targ[style={draw=magenta}]{} 	& \targ{}				& \ctrl{-2}	& \ctrl{-1}	& \rstick{$\ket{a_1}$} \\
	\lstick{$\ket{b_2}$}	& \targ{}	& 			& 			& 			& \ctrl[style={draw=orange,fill=orange}]{1} 	&			& 			&			&			&			& \ctrlo[style={draw=magenta,fill=magenta}]{1} & 			& 						& 			& \targ{}	& \rstick{$\ket{s_2}$} \\
	\lstick{$\ket{a_2}$}	& \ctrl{-1}	& 			& \ctrl{-2}	& \targ{}	& \targ[style={draw=orange}]{} 	& \ctrl[style={draw=orange,fill=orange}]{1}	&			&			&			& \ctrl[style={draw=magenta,fill=magenta}]{1}	& \targ[style={draw=magenta}]{} 	& \targ{}	& \ctrl{-2}				& 			& \ctrl{-1}	& \rstick{$\ket{a_2}$} \\
	\lstick{$\ket{b_3}$}	& \targ{}	& 			& 			& 			&			& \ctrl[style={draw=orange,fill=orange}]{1} 	&			& 			&			& \ctrlo[style={draw=magenta,fill=magenta}]{1} & 			& 			& 						&			& \targ{}	& \rstick{$\ket{s_3}$} \\
	\lstick{$\ket{a_3}$}	& \ctrl{-1}	& 			&			& \ctrl{-2}	& \targ{}	& \targ[style={draw=orange}]{} 	& \ctrl[style={draw=orange,fill=orange}]{1}	&			& \ctrl[style={draw=magenta,fill=magenta}]{1}	& \targ[style={draw=magenta}]{} 	& \targ{}	& \ctrl{-2}	& 						&			& \ctrl{-1}	& \rstick{$\ket{a_3}$} \\
	\lstick{$\ket{b_4}$}	& \targ{}	& 			&			& 			& 			& 			& \ctrl[style={draw=orange,fill=orange}]{1}	&			& \ctrlo[style={draw=magenta,fill=magenta}]{1} & 			& 			& 			&						&			& \targ{}	& \rstick{$\ket{s_4}$} \\
	\lstick{$\ket{a_4}$}	& \ctrl{-1}	& 			&			& 			& \ctrl{-2}	& \targ{}	& \targ[style={draw=orange}]{}	& \ctrl{1}	& \targ[style={draw=magenta}]{} 	& \targ{}	& \ctrl{-2}	& 			& 						&			& \ctrl{-1}	& \rstick{$\ket{a_4}$} \\
	\lstick{$\ket{b_5}$}	& \targ{}	& 			&			& 			& 			& 			&			& \ctrlo{2} & 			& 			& 			& 			&						&			& \targ{}	& \rstick{$\ket{s_5}$} \\
	\lstick{$\ket{a_5}$}	& \ctrl{-1}	& 			&			& 			& 			& \ctrl{-2}	& \ctrl{1}	& 			& 			& \ctrl{-2}	& 			& 			&						&			& \ctrl{-1}	& \rstick{$\ket{a_5}$} \\
	\lstick{$\ket{z}$}		& 			& 			&			& 			& 			&			& \targ{}	& \targ{}	& 			& 			& 			& 			&						&			& 			& \rstick{$\ket{z \oplus s_6}$}
	\end{quantikz}}
	\caption{Circuit from \cite{CDKM04} (Figure 6, $\ie$, Figure 4 combined with Figure 2(b)) for the addition of numbers of length 6. }\label{circ:Cuccaro}
\end{figure}

\subsection{Size-Efficient Ripple-Carry Adder}\label{sec:Cheap}

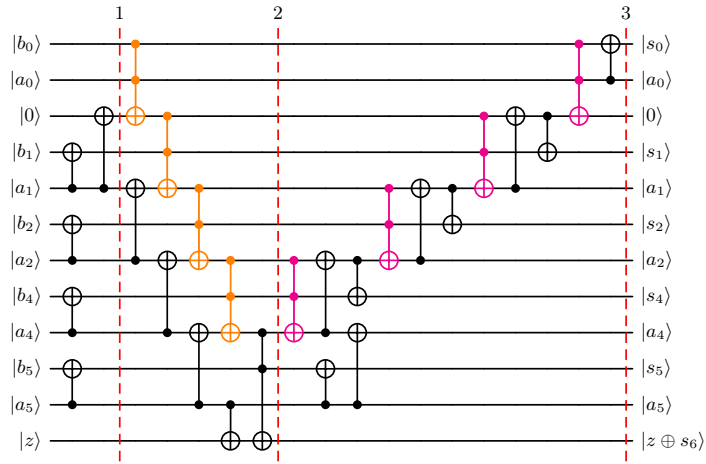
\begin{figure}[!ht]\centering
	\scalebox{0.8}{
	\begin{quantikz}[column sep=.2cm, row sep={.6cm,between origins}]
	\lstick{$\ket{b_0}$}	& 			& \slice{1}	& \ctrl[style={draw=orange,fill=orange}]{1} 	&			&			&			& \slice{2}	&			&			&			& 			&			&			& 			& 			& 			& \ctrl[style={draw=magenta,fill=magenta}]{1} & \targ{} \slice{3}	& \rstick{$\ket{s_0}$} \\
	\lstick{$\ket{a_0}$}	& 			& 			& \ctrl[style={draw=orange,fill=orange}]{1}	& 			& 			&			& 			&			&			&			& 			&			&			& 			& 			& 			& \ctrl[style={draw=magenta,fill=magenta}]{1}		& \ctrl{-1}		& \rstick{$\ket{a_0}$} \\
	\lstick{$\ket{0}$}		&			& \targ{}	& \targ[style={draw=orange}]{}	& \ctrl[style={draw=orange,fill=orange}]{1}	&			&			& 			&			&			&			&			&			&			& \ctrl[style={draw=magenta,fill=magenta}]{1}	& \targ{}	& \ctrl{1}& \targ[style={draw=magenta}]{}			&& \rstick{$\ket{0}$} \\
	\lstick{$\ket{b_1}$}	& \targ{}	& 			& 			& \ctrl[style={draw=orange,fill=orange}]{1} & 			&			& 			&			&			&			&			&			&			& \ctrl[style={draw=magenta,fill=magenta}]{1}	& 			& \targ{}	& 						&& \rstick{$\ket{s_1}$} \\
	\lstick{$\ket{a_1}$}	& \ctrl{-1}	& \ctrl{-2}	& \targ{}	& \targ[style={draw=orange}]{} 	& \ctrl[style={draw=orange,fill=orange}]{1}	&			& 			&			&			&			& \ctrl[style={draw=magenta,fill=magenta}]{1}	& \targ{}	& \ctrl{1}	& \targ[style={draw=magenta}]{} 	& \ctrl{-2}	& 			&						&& \rstick{$\ket{a_1}$} \\
	\lstick{$\ket{b_2}$}	& \targ{}	& 			& 			& 			& \ctrl[style={draw=orange,fill=orange}]{1} &			& 			&			&			&			& \ctrl[style={draw=magenta,fill=magenta}]{1} & 			& \targ{}	& 			& 			& 			& 						&& \rstick{$\ket{s_2}$} \\
	\lstick{$\ket{a_2}$}	& \ctrl{-1}	& 			& \ctrl{-2}	& \targ{}	& \targ[style={draw=orange}]{} 	& \ctrl[style={draw=orange,fill=orange}]{1}	&			& \ctrl[style={draw=magenta,fill=magenta}]{1}	& \targ{}	& \ctrl{1}	& \targ[style={draw=magenta}]{} 	& \ctrl{-2}	& 			& 			& 			& 			& 						&& \rstick{$\ket{a_2}$} \\
	\lstick{$\ket{b_4}$}	& \targ{}	& 			&			& 			& 			& \ctrl[style={draw=orange,fill=orange}]{1}	&			& \ctrl[style={draw=magenta,fill=magenta}]{1} &			& \targ{}	& 			&			& 			& 			& 			& 			& 						&& \rstick{$\ket{s_4}$} \\
	\lstick{$\ket{a_4}$}	& \ctrl{-1}	& 			&			& \ctrl{-2}	& \targ{}	& \targ[style={draw=orange}]{}	& \ctrl{1}	& \targ[style={draw=magenta}]{} 	& \ctrl{-2}	& \targ{}	& 			&			& 			& 			& 			& 			& 						&& \rstick{$\ket{a_4}$} \\
	\lstick{$\ket{b_5}$}	& \targ{}	& 			&			& 			& 			&			& \ctrl{2} &			& \targ{}	& 			& 			&			& 			& 			& 			& 			& 						&& \rstick{$\ket{s_5}$} \\
	\lstick{$\ket{a_5}$}	& \ctrl{-1}	& 			&			& 			& \ctrl{-2}	& \ctrl{1}	& 			&			& \ctrl{-1}	& \ctrl{-2}	& 			&			& 			& 			& 			& 			& 						&& \rstick{$\ket{a_5}$} \\
	\lstick{$\ket{z}$}		& 			& 			&			& 			&			& \targ{}	& \targ{}	& 			& 			& 			&			&			& 			& 			& 			& 			& 						&& \rstick{$\ket{z \oplus s_6}$}
	\end{quantikz}}
	\caption{Circuit from \cite{CDKM04} (Figure 4 combined with Figure 2(a)) for the addition of numbers of length 6. }\label{circ:CuccaroV2}
\end{figure}

We continue with the second circuit from \cite{CDKM04}, namely their Figure 4 combined with Figure 2(a). An example for $n=6$ is provided in \autoref{circ:CuccaroV2}. A V-shaped circuit can be find in there, with a left branch of Toffoli gates and a right branch of Toffoli gates interspersed with $\CNOT$ gates.

Without any optimization, it can be verified that for any $n$, Slice 1 has a $\CNOT$-depth of 2 and $\CNOT$-count of $n-1$. Slice 2 has a $\CNOT$-depth of $7(n-1)$, a $\CNOT$-count of $8n-9$, a $\T$-depth of $3(n-1)$ and a $\T$-count of $7(n-1)$. Slice 3 has a $\CNOT$-depth of $9n-19$, a $\CNOT$-count of $9n-17$, a $\T$-depth of $3(n-2)$ and a $\T$-count of $7(n-2)$. In total, a $\CNOT$-depth of $16n-24$, $\CNOT$-count of $18n-17$, $\T$-depth of $6n-9$ and $\T$-count of $14n-21$ are exepected.

Now, taking back the metrics given in \autoref{sec:ToffToff} for a V-shaped circuit with Toffoli gates on the left branch and Toffoli gates interspersed with two $\CNOT$ gates on the right one (with $n-1$ layers): we have for the optimized Cuccaro $\etal$'s adder a $\T$-depth of $4(n-1)$, a $\T$-count of $10(n-1)$, a $\CNOT$-depth of $11n-10$ ($4n-3$ for the left branch and $7(n-1)$ for the right one after simplification of the matching gates and because of the interspersed $\CNOT$ gates) and a $\CNOT$-count of $12(n-1)$ ($7(n-1)$ for the left branch and $9(n-1)$ for the right one, minus $4(n-1)$ for the matching gates). The operator $\U$ is a Peres gate. Using \autoref{circ:PeresDec}, we have to add 3 to the $\CNOT$-depth and 2 to the $\T$-depth (the majority of the gates being parallelizable with gates of the V-shape), 5 to the $\CNOT$-count and 7 to the $\T$-count. For the rest of the circuit, we have $2n-1$ additionnal $\CNOT$ gates, but the majority can once again be computed in parallel with the V-shape, thus we have an additionnal small increase of 2 for the $\CNOT$-depth. In total, the circuit in \autoref{circ:Cuccaro} has $\T$-depth of $4n-2$, $\T$-count of $10n-3$, $\CNOT$-depth of $11n-8$ and $\CNOT$-count of $14n-10$.

\subsection{Ancilla-Free Ripple-Carry Adder}\label{sec:AncFree}

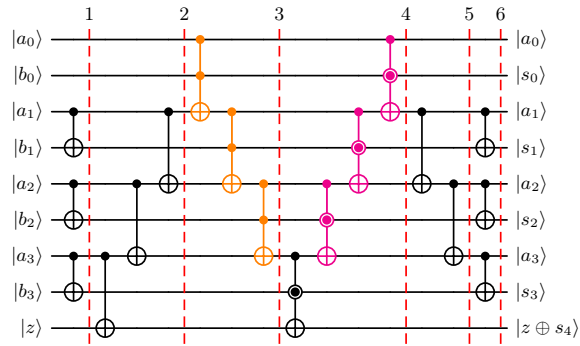
\begin{figure}[!ht]\centering
	\scalebox{0.8}{
	\begin{quantikz}[column sep=.2cm, row sep={.6cm,between origins}]
	\lstick{$\ket{a_0}$}	& \slice{1}	& 			& 			& \slice{2}	& \ctrl[style={draw=orange,fill=orange}]{1}	& 			& \slice{3}	&	 		& 			& 			& \ctrl[style={draw=magenta,fill=magenta}]{1} \slice{4}	&	& \slice{5}	& \slice{6}	& \rstick{$\ket{a_0}$} \\
	\lstick{$\ket{b_0}$}	&			& 			& 			& 			& \ctrl[style={draw=orange,fill=orange}]{1} 	& 			& 			& 			& 			& 			& \ctrlo[style={draw=magenta,fill=magenta}]{1}	& 			& 			& 			& \rstick{$\ket{s_0}$} \\
	\lstick{$\ket{a_1}$}	& \ctrl{1}	& 			& 			& \ctrl{2}	& \targ[style={draw=orange}]{}	& \ctrl[style={draw=orange,fill=orange}]{1}	&			& 			& 			& \ctrl[style={draw=magenta,fill=magenta}]{1} 	& \targ[style={draw=magenta}]{}	& \ctrl{2}	& 			& \ctrl{1}	& \rstick{$\ket{a_1}$} \\
	\lstick{$\ket{b_1}$}	& \targ{}	& 			& 			& 			& 			& \ctrl[style={draw=orange,fill=orange}]{1} 	&			& 			& 			& \ctrlo[style={draw=magenta,fill=magenta}]{1}	& 			& 			& 			& \targ{}	& \rstick{$\ket{s_1}$} \\
	\lstick{$\ket{a_2}$}	& \ctrl{1}	&			& \ctrl{2}	& \targ{}	& 			& \targ[style={draw=orange}]{}	& \ctrl[style={draw=orange,fill=orange}]{1} 	& 			& \ctrl[style={draw=magenta,fill=magenta}]{1} 	& \targ[style={draw=magenta}]{}	&			& \targ{}	& \ctrl{2}	& \ctrl{1}	& \rstick{$\ket{a_2}$} \\
	\lstick{$\ket{b_2}$}	& \targ{}	& 			& 			& 			& 			& 			& \ctrl[style={draw=orange,fill=orange}]{1} 	& 			& \ctrlo[style={draw=magenta,fill=magenta}]{1}	& 			&			& 			& 			& \targ{}	& \rstick{$\ket{s_2}$} \\
	\lstick{$\ket{a_3}$}	& \ctrl{1}	& \ctrl{2}	& \targ{}	& 			& 			& 			& \targ[style={draw=orange}]{}	& \ctrl{1}	& \targ[style={draw=magenta}]{}	&			&			&			& \targ{}	& \ctrl{1}	& \rstick{$\ket{a_3}$} \\
	\lstick{$\ket{b_3}$}	& \targ{}	& 			& 			& 			& 			& 			& 			& \ctrlo{1}	& 			& 			&			& 			& 			& \targ{}	& \rstick{$\ket{s_3}$} \\
	\lstick{$\ket{z}$} 	& 			& \targ{}	& 			&			& 			& 			&			& \targ{}	& 			& 			& 			&			&			&			& \rstick{$\ket{z \oplus s_4}$}
	\end{quantikz}}
	\caption{Ancilla-free adder represented as a circuit for $n=4$, proposed in \cite{TTK10}.}\label{circ:TTK}
\end{figure}

We end with the circuit from \cite{TTK10}, (also studied in \cite{TR13}). An example for $n=4$ is provided in \autoref{circ:TTK}. A V-shaped circuit can be find in there, with a left branch of Toffoli gates and a right branch of Peres gates.

Without any optimization, it can be verified that for any $n$, Slices 1 and 6 of their circuit have a $\CNOT$-depth of 1 and $\CNOT$-count of $n-1$. Slice 2 has a $\CNOT$-depth and $\CNOT$-count of $n-1$, Slice 5 of $n-2$. Slice 3 has a $\CNOT$-depth and $\CNOT$-count of $7(n-1)$, a $\T$-depth of $3(n-1)$ and a $\T$-count of $7(n-1)$. Slice 4 has a $\CNOT$-depth and $\CNOT$-count of $6n$, a $\T$-depth of $3n$ and a $\T$-count of $7n$. In total, a $\CNOT$-depth of $15n-8$, $\CNOT$-count of $17n-12$, $\T$-depth of $6n-3$ and $\T$-count of $14n-7$ are exepected.

Now, taking back the metrics given in \autoref{sec:ToffPer} for a V-shaped circuit with Toffoli gates on the left branch and Peres gates on the right one (with $n-1$ layers): we have for the optimized Takahashi $\etal$'s adder a $\T$-depth of $3n+1$, a $\T$-count of $12n-12$, a $\CNOT$-depth of $8n-4$ and a $\CNOT$-count of $12n-12$. The operator $\U$ is a Peres gate. Using \autoref{circ:PeresDec}, we have to add 2 to the $\CNOT$-depth and 1 to the $\T$-depth (the majority of the gates being parallelizable with gates of the V-shape), 5 to the $\CNOT$-count and 7 to the $\T$-count. For the rest of the circuit, we have $4n-5$ additionnal $\CNOT$ gates, but the majority can once again be computed in parallel, thus we have an additionnal increase of $2n-1$ for the $\CNOT$-depth. In total, the circuit in \autoref{circ:TTK} has $\T$-depth of $3n+2$, $\T$-count of $12n-5$, $\CNOT$-depth of $10n-3$ and $\CNOT$-count of $16n-12$.

\section{Optimizing Ripple-Carry Comparators}

We proceed in three stages to compare $a$ and $b$, using the strategy described in \cite{CDKM04}. In the first one, we simply apply $\X$ gates on every single bit of $b$ and then proceed as in the adders with the left-hand branch of the V-shape. In the second stage, we apply a $\CNOT$ with control qubit $A_{n-1}$ and target qubit $Z$. Finally, in the third stage, we uncompute the first one.

We take up the adders from \autoref{sec:Shallow} and \autoref{sec:AncFree} to present their optimized versions. Note that the adder in \autoref{sec:Cheap} matches the one in \autoref{sec:Shallow} on the left-hand side of the V-shaped circuit, so that both adders produce the same comparator by its mirroring definition.

Since studying the complexity of these circuits without optimization is very similar to what we have done for the corresponding adders, we just refer to \autoref{table:Comp} for the values.

\subsection{Shallow Ripple-Carry Comparator}\label{sec:CompCDKM}

\begin{figure}[!ht]\centering
	\scalebox{0.8}{
	\begin{quantikz}[column sep=.2cm, row sep={.6cm,between origins}]
	\lstick{$\ket{b_0}$}& 			& \gate{\X} \slice{1}& \ctrl[style={draw=orange,fill=orange}]{1}	& 	& \slice{2}	& \slice{3}	&		 	& 			& \ctrl[style={draw=magenta,fill=magenta}]{1} \slice{4}& \gate{\X}& \slice{5} & \rstick{$\ket{b_0}$} \\
	\lstick{$\ket{a_0}$}& 			& 			& \ctrl[style={draw=orange,fill=orange}]{1} & 			&			& 			& 			& 			& \ctrl[style={draw=magenta,fill=magenta}]{1}	& 			&			& \rstick{$\ket{a_0}$} \\
	\lstick{$\ket{0}$}	& 			& \targ{}	& \targ[style={draw=orange}]{} 	& \ctrl[style={draw=orange,fill=orange}]{1}	&			& 			& 			& \ctrl[style={draw=magenta,fill=magenta}]{1}	& \targ[style={draw=magenta}]{} 	& \targ{}	&			& \rstick{$\ket{0}$} \\
	\lstick{$\ket{b_1}$}& \targ{}	& 			& \gate{\X}	& \ctrl[style={draw=orange,fill=orange}]{1} &			& 			& 			& \ctrl[style={draw=magenta,fill=magenta}]{1} & \gate{\X}	& 			& \targ{}	& \rstick{$\ket{b_1}$} \\
	\lstick{$\ket{a_1}$}& \ctrl{-1}	& \ctrl{-2}	& \targ{}	& \targ[style={draw=orange}]{} 	& \ctrl[style={draw=orange,fill=orange}]{1}	& 			& \ctrl[style={draw=magenta,fill=magenta}]{1}	& \targ[style={draw=magenta}]{} 	& \targ{}	& \ctrl{-2}	& \ctrl{-1}	& \rstick{$\ket{a_1}$} \\
	\lstick{$\ket{b_2}$}& \targ{}	& 			& 			& \gate{\X}	& \ctrl[style={draw=orange,fill=orange}]{1}	& 			& \ctrl[style={draw=magenta,fill=magenta}]{1} & \gate{\X}	& 			& 			& \targ{}	& \rstick{$\ket{b_2}$} \\
	\lstick{$\ket{a_2}$}& \ctrl{-1}	& 			& \ctrl{-2}	& \targ{}	& \targ[style={draw=orange}]{}	& \ctrl{1}	& \targ[style={draw=magenta}]{}	& \targ{}	& \ctrl{-2}	& 			& \ctrl{-1}	& \rstick{$\ket{a_2}$} \\
	\lstick{$\ket{b_3}$}& \targ{}	& 			& 			& 			& \gate{\X}	& \ctrl{2}	& \gate{\X}	& 			& 			& 			& \targ{}	& \rstick{$\ket{b_3}$} \\
	\lstick{$\ket{a_3}$}& \ctrl{-1}	& 			& 			& \ctrl{-2}	& \ctrl{1}	&			& 			& \ctrl{-2}	& 			& 			& \ctrl{-1}	& \rstick{$\ket{a_3}$} \\
	\lstick{$\ket{z}$}	& 			& 			& 			& 			& \targ{}	& \targ{}	& 			& 			& 			& 			& 			& \rstick{$\ket{z \oplus (a \le b)}$}
	\end{quantikz}}
	\caption{Comparator derived from \autoref{circ:Cuccaro}, given in \cite{CDKM04}, represented as a circuit for $n=4$.} \label{circ:CompCuccaro}
\end{figure}
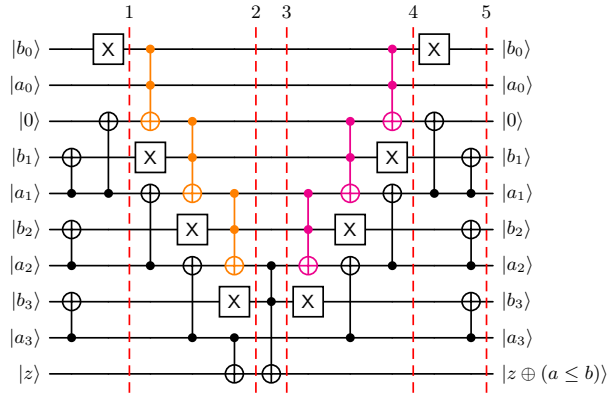

We take back the adder from \autoref{sec:Shallow} and turn it into a comparator. An example for $n=4$ is provided in \autoref{circ:CompCuccaro}.

Taking back the metrics given in \autoref{sec:ToffToff}, we have for the V-shape of the optimized comparator a $\T$-depth of $4n-4$, a $\T$-count of $10n-10$, a $\CNOT$-depth of $8n-6$ and a $\CNOT$-count of $10n-10$. The operator $\U$ is a Peres gate. Using \autoref{circ:PeresDec}, we have to add 7 to the $\CNOT$-depth and 3 to the $\T$-depth, 7 to the $\CNOT$-count and 7 to the $\T$-count. For the rest of the circuit, we have $4n-3$ additionnal $\CNOT$ gates, but the majority can be computed in parallel with the V-shape, thus we have an additionnal small increase of 4 for the $\CNOT$-depth. In total, the circuit in \autoref{circ:CompCuccaro} has $\T$-depth of $4n-1$, $\T$-count of $10n-3$, $\CNOT$-depth of $8n+5$ and $\CNOT$-count of $14n-6$.

\subsection{Ancilla-Free Ripple-Carry Comparator}\label{sec:CompTTK}

\begin{figure}[!ht]\centering
	\scalebox{0.8}{
	\begin{quantikz}[column sep=.2cm, row sep={.6cm,between origins}]
	\lstick{$\ket{a_0}$}	& \slice{1}	& 			& 			& \slice{2}	& \ctrl[style={draw=orange,fill=orange}]{1}	& 			& \slice{3}	& \slice{4}	& 			& 			& \ctrl[style={draw=magenta,fill=magenta}]{1} \slice{5} &	& \slice{6}	& \slice{7}	& \rstick{$\ket{a_0}$} \\
	\lstick{$\ket{b_0}$}	&			& 			& 			& \gate{\X}	& \ctrl[style={draw=orange,fill=orange}]{1} 	& 			& 			& 			& 			& 			& \ctrl[style={draw=magenta,fill=magenta}]{1}	& \gate{\X}	& 			& 			& \rstick{$\ket{b_0}$} \\
	\lstick{$\ket{a_1}$}	& \ctrl{1}	& 			& 			& \ctrl{2}	& \targ[style={draw=orange}]{}	& \ctrl[style={draw=orange,fill=orange}]{1}	&			& 			& 			& \ctrl[style={draw=magenta,fill=magenta}]{1} 	& \targ[style={draw=magenta}]{}	& \ctrl{2}	& 			& \ctrl{1}	& \rstick{$\ket{a_1}$} \\
	\lstick{$\ket{b_1}$}	& \targ{}	& 			& 			& 			& \gate{\X}	& \ctrl[style={draw=orange,fill=orange}]{1} 	&			& 			& 			& \ctrl[style={draw=magenta,fill=magenta}]{1}	& \gate{\X}	& 			& 			& \targ{}	& \rstick{$\ket{b_1}$} \\
	\lstick{$\ket{a_2}$}	& \ctrl{1}	&			& \ctrl{2}	& \targ{}	& 			& \targ[style={draw=orange}]{}	& \ctrl[style={draw=orange,fill=orange}]{1} 	& 			& \ctrl[style={draw=magenta,fill=magenta}]{1} 	& \targ[style={draw=magenta}]{}	&			& \targ{}	& \ctrl{2}	& \ctrl{1}	& \rstick{$\ket{a_2}$} \\
	\lstick{$\ket{b_2}$}	& \targ{}	& 			& 			& 			& 			& \gate{\X}	& \ctrl[style={draw=orange,fill=orange}]{1} 	& 			& \ctrl[style={draw=magenta,fill=magenta}]{1}	& \gate{\X}	&			& 			& 			& \targ{}	& \rstick{$\ket{b_2}$} \\
	\lstick{$\ket{a_3}$}	& \ctrl{1}	& \ctrl{2}	& \targ{}	& 			& 			& 			& \targ[style={draw=orange}]{}	& \ctrl{1}	& \targ[style={draw=magenta}]{}	&			&			&			& \targ{}	& \ctrl{1}	& \rstick{$\ket{a_3}$} \\
	\lstick{$\ket{b_3}$}	& \targ{}	& 			& 			& 			& 			& 			& 			& \ctrlo{1}	& 			& 			&			& 			& 			& \targ{}	& \rstick{$\ket{b_3}$} \\
	\lstick{$\ket{z}$} 	& 			& \targ{}	& 			&			& 			& 			&			& \targo{}	& 			& 			& 			&			&			&			& \rstick{$\ket{z \oplus (a \le b)}$}
	\end{quantikz}}
	\caption{Ancilla-free comparator derived from \autoref{circ:TTK}, given in \cite{TTK10}, represented as a circuit for $n=4$}\label{circ:CompTTK}
\end{figure}
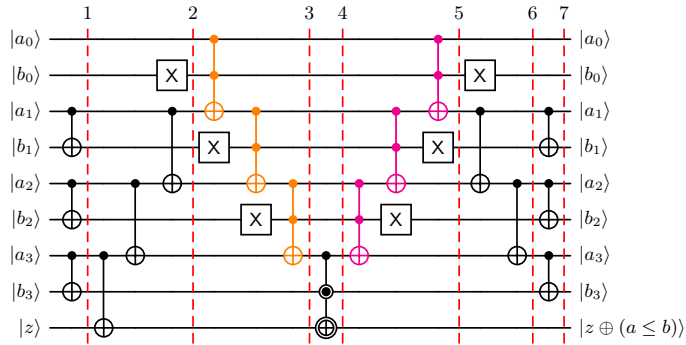

We take back the adder from \autoref{sec:AncFree} and turn it into a comparator. An example for $n=4$ is provided in \autoref{circ:CompTTK}.

Once again, taking back the metrics given in \autoref{sec:ToffToff}, we have for the optimized comparator a $\T$-depth of $4n-4$, a $\T$-count of $10n-10$, a $\CNOT$-depth of $8n-6$ and a $\CNOT$-count of $10n-10$. The operator $\U$ is a TR gate. Using \autoref{circ:PeresDec}, we have to add 2 to the $\CNOT$-depth and 1 to the $\T$-depth (the majority of the gates being parallelizable with gates of the V-shape), 6 to the $\CNOT$-count and 7 to the $\T$-count. Now for the rest of the circuit, we have $4n-5$ additionnal $\CNOT$ gates, but several can be computed in parallel, thus we have an additionnal increase of $2n-1$ for the $\CNOT$-depth. In total, the circuit in \autoref{circ:CompTTK} has $\T$-depth of $4n-3$, $\T$-count of $10n-3$, $\CNOT$-depth of $10n-6$ and $\CNOT$-count of $14n-9$.

\section{Conclusion}

We have shown that Cuccaro $\etal$'s adders and Takahashi $\etal$'s adder do not actually have a $\T$-depth in $6n + O(1)$, but in fact in $3n + O(1)$, without resorting to tricks involving measurements or approximate gates. We also studied their $\CNOT$-count and $\CNOT$-depth, and did the same for the corresponding comparators. On another front, we showed that a $\T$-count in $10n+O(1)$ and a $\CNOT$-count in $14n+O(1)$ are achievable.

An interesting question is whether it is possible to go further, either in terms of count or depth, and $\T$ gates or $\CNOT$ gates. Another question would be whether these optimized circuits can give better results than what is already known when using techniques with measurements \cite{Gid18} or approximate gates \cite{LFXL22}.

\subsection*{Acknowledgments}

This work is part of HQI initiative (www.hqi.fr) and is supported by France 2030 under the French National Research Agency award number “ANR-22-PNCQ-0002”.

\newcommand{\etalchar}[1]{$^{#1}$}

\end{document}